\documentclass{ws-procs9x6-cpt19}

\begin{document}

\newcommand{\refeq}[1]{(\ref{#1})}
\def\etal {{\it et al.}}

\title{Lorentz Violation and Radiative Corrections in Gauge Theories}

\author{A.F.\ Ferrari}

\address{Centro de Ci\^encias Naturais e Humanas, Federal do ABC,\\ 
Avenida dos Estados, 5001, Santo Andr\'e - SP ,09210-580, Brazil
}

\begin{abstract}
Various studies have already considered 
radiative corrections in Lorentz-violating models 
unveiling many instances where a minimal or nonminimal operator generates, 
via loop corrections, 
a contribution to the photon sector of the Standard-Model Extension. 
However, 
an important fraction of this literature 
does not follow the widely accepted conventions and notations
of the Standard-Model Extension, 
and this obscures the comparison 
between different calculations 
as well as possible phenomenological consequences. 
After reviewing some of these works, 
we uncover one example 
where a well defined loop correction to the $k_{F}$ coefficient, 
already presented in the literature, 
allows us to improve the bounds on one specific coefficient 
of the fermion sector of the Lorentz-violating QED extension.
\end{abstract}

\bodymatter

\phantom{}\vskip10pt\noindent

The Standard-Model Extension (SME)\cite{coll1,coll2} 
is understood as an effective field theory 
based on the internal symmetries and field content of the Standard Model, 
but incorporating Lorentz violation (LV) 
in a very general way. 
A broad experimental program has used the SME framework 
to obtain stringent bounds on possible LV operators 
using different experiments and astrophysical observations.\cite{datatables}
The most studied and well constrained sector of the SME 
is the LV extension to the Maxwell theory, 
defined in its most general form in Ref.\,\refcite{KostelMewes-EMHD}. 
In particular, 
the minimal photon sector involves only two LV coefficients,
 $k_{AF}$ and $k_{F}$.
Both terms in general induce birefringence in the vacuum 
leading to very strong constraints from astrophysics: 
of order $10^{-43}\,$GeV for $k_{AF}$ 
and $10^{-37}$ for the birefringent components of $k_{F}$.\cite{datatables}

If taken as an effective parametrization for LV effects 
to be searched for in the laboratory, 
the SME might be understood as a strictly tree-level theory. 
From the theoretical viewpoint, 
however, 
it motivates many interesting studies 
concerning the consistency, 
as a full quantum field theory (QFT), 
of theories in which one of the central aspects of the QFT formalism---Lorentz
symmetry---is in some sense violated.
Different field-theoretical aspects, 
such as renormalization, 
the structure of asymptotic states, 
and others, 
have already been investigated.\cite{KostPic,Colladay:2009rb,Ferrero:2011yu,Potting:2011yj,Cambiaso:2014eba}

Radiative corrections, 
in particular, 
may lead to results of direct phenomenological interest: 
some set of LV coefficients 
may generate or contribute to other sets of LV coefficients 
via loop corrections.
An early example was discussed in detail in Ref.\ \refcite{JK}:
the $b$ term
\begin{equation}
b_{\mu}\bar{\psi}\gamma^{\mu}\gamma_{5}\psi,\label{eq:Vb}
\end{equation}
which is a LV correction to the fermion propagator, 
will induce a one-loop correction to the quadratic photon lagrangian 
of the form
\begin{equation}
Ce^{2}\epsilon^{\mu\nu\lambda\rho}b_{\mu}\thinspace A_{\nu}\partial_{\lambda}A_{\rho},\label{cfj}
\end{equation}
$e$ being the fermion charge and $C$ a constant. 
In the SME notation, 
this amounts to the generation of a minimal $k_{AF}$ term
with $k_{AF}\sim b$ 
or to a correction to a $k_{AF}$ term
already present at tree level. 
The phenomenological interest in such a result
is that $k_{AF}$ can be strongly constrained 
by photon vacuum birefringence,
and these constraints could be translated to $b^{\mu}$. 
However,
it was readily recognized 
that the result presented in Eq.\ \eqref{cfj} is anomalous: 
the loop integral turns out to be finite but ambiguous,
its result being dependent on the regularization scheme 
used to calculate it 
(for a recent discussion of this problem, 
see Ref.\ \refcite{Altschul:2019eip}).
In this case, 
it is certainly not possible 
to use loop corrections to obtain sound phenomenological conclusions.

Many different instances of radiative generation of LV operators 
have been presented in the literature, 
and we might wonder whether finite and well-defined corrections 
can be calculated in some cases,
and whether stringent bounds obtained in one sector of the SME 
might be transferred to other, 
perhaps not so well bounded sectors. 
This requires some work 
since many of the reported calculations 
do not follow the now standard SME notations.

As an example, 
higher-derivative corrections to the photon sector
originating from \refeq{eq:Vb} 
where presented in Ref.\ \refcite{TMariz1} as 
\begin{equation}
{\cal L}_{eff}\supset\frac{e^{2}}{24\pi^{2}m^{2}}\thinspace\epsilon^{\beta\mu\nu\rho}b_{\beta}\thinspace A_{\mu}\Box F_{\nu\rho},\label{HDCFJ}
\end{equation}
corresponding in SME notation to the generation of the dimension five coefficient 
\begin{equation}
(\hat{k}_{AF}^{\left(5\right)})_{\kappa}^{\hphantom{\kappa}\alpha\beta}=\frac{e^{2}}{48\pi^{2}m^{2}}b_{\kappa}\eta^{\alpha\beta}.\label{eq:HDCFJkAF}
\end{equation}
This is a finite result, 
free of ambiguities. 
Unfortunately, 
this form of $\hat{k}_{AF}^{\left(5\right)}$ 
does not modify free propagation of photons 
at leading order, 
so no interesting bounds can be obtained
from this result at the moment.

Looking at one-loop corrections 
involving higher orders in $b$, 
we may also obtain finite and well-defined results, 
such as\cite{Bonneau,aether2}
\begin{equation}
{\cal L}_{eff}\supset-\frac{e^{2}}{6\pi^{2}m^{2}}\thinspace b^{\mu}b_{\lambda}\thinspace F_{\mu\nu}F^{\lambda\nu},\label{aether}
\end{equation}
which, 
written in the SME notation, 
corresponds to
\begin{equation}
(k_{F}^{(4)})^{\mu\nu\alpha\beta}=-\frac{e^{2}}{6\pi^{2}m^{2}}\left(b^{\mu}b^{\alpha}\eta^{\nu\beta}-b^{\nu}b^{\alpha}\eta^{\mu\beta}-b^{\mu}b^{\beta}\eta^{\nu\alpha}+b^{\nu}b^{\beta}\eta^{\mu\alpha}\right).\label{eq:aetherkF}
\end{equation}
At first sight, 
one might not expect to find competitive constraints
from this expression, 
since it is of second order in $b$. 
However,
birefringent components of $k_{F}$ are constrained at the order of $10^{-37}$,
and so this result would provide a bound of the order 
$b^{2}<6\pi^{2}m^{2}/e^{2}\times10^{-37}$
for the $b$ coefficient, 
corresponding to $\left|b_{p}^{\mu}\right|<3\times10^{-15}\,$GeV
for protons 
and $\left|b_{e}^{\mu}\right|<1.5\times10^{-20}\,$GeV
for electrons, 
for example. 
These are not better 
(but also not much worse) 
than the constraints already found 
{\it for the space components} of $b^{\mu}$, 
as measured in the Sun-centered reference frame.\cite{datatables}
On the other hand, 
the temporal component $b^{T}$ 
is not so well constrained: 
the best bounds are of order $\left|(b^{T})_{p}\right|<7\times10^{-8}\,$GeV
for protons 
and $\left|(b^{T})_{e}\right|<10^{-15}\,$GeV
for electrons. 
Therefore, 
the radiative correction presented in Eq.\ \eqref{eq:aetherkF}
can translate the stringent constraints on birefringent components of $k_{F}$ 
to competitive bounds on the temporal components $b^{T}$.
It remains to check 
that this particular form of $k_{F}^{(4)}$ 
does indeed induce birefringence. 
The easiest way to do this 
is by using the parametrization of birefringent components of $k_{F}$ 
in terms of the ten $k^{a}$ coefficients 
as given in Ref.\ \refcite{Kostelecky:2001mb}:
one may easily verify that Eq.\ \eqref{eq:aetherkF} 
corresponds to non-vanishing coefficients $k^{a}=e^{2}b^{2}/6\pi^{2}m^{2}$ 
for $a=3,4$.

The end result is therefore the constraint
\begin{equation}
\left|b_{T}\right|<\pi m/e\sqrt{6\times10^{-37}}\thinspace,
\end{equation}
for the temporal component, 
in the Sun-centered frame, 
of the $b$ coefficient for a given fermion, 
depending on its mass $m$.
For example, 
we have 
\begin{equation}
\left|(b_{T})_{p}\right|<3\times10^{-15}\thinspace\text{GeV}
\end{equation}
for protons and 
\begin{equation}
\left|(b_{T})_{e}\right|<1.5\times10^{-20}\thinspace\text{GeV}
\end{equation}
for electrons.

This result was presented in Ref.\ \refcite{Ferrari:2018tps}, 
together with an extensive study of other instances of radiative corrections
in different sectors of the SME. 
It is an interesting example, 
where a weakly bounded coefficient for LV 
can be subjected to a stronger constraint, 
borrowed from the very well studied photon sector of the SME. 
We believe that, 
besides interesting questions regarding theoretical consistency 
and technical challenges of calculating ambiguous Feynman integrals, 
the study of radiative corrections
might help to fill some of the gaps 
in the extensive set of searches for LV 
that have been developed in the last decades 
using the SME as the fundamental framework.

\section*{Acknowledgments}
This work was partially supported 
by Funda\c c\~ao de Amparo \`a Pesquisa do Estado de S\~ao Paulo (FAPESP) 
and Conselho Nacional de Desenvolvimento Cient\'\i fico e Tecnol\'ogico (CNPq) 
via the following grants: CNPq 304134/2017-1 and FAPESP 2017/13767-9.

\end{document}